\documentclass[aps,prd,showpacs,superscriptaddress,nofootinbib,10pt]{revtex4-2}
\usepackage{acronym}
\usepackage{amsfonts}
\usepackage{amsmath}

\usepackage{amssymb}
\usepackage{bm}
\usepackage{cases}
\usepackage{color}
\usepackage{comment}
\usepackage[dvipsnames]{xcolor}
\usepackage{enumerate}
\usepackage{epsfig}
\usepackage{etoolbox}
\usepackage{fancyref}
\usepackage{fancyhdr}
\usepackage[T1]{fontenc} 
\usepackage{graphicx}
\usepackage{hyperref}
\usepackage{indentfirst}
\usepackage{latexsym}
\usepackage{mathrsfs}
\usepackage{mciteplus}
\usepackage{multirow}
\usepackage{rotating}
\usepackage{scrextend}
\usepackage{soul}
\usepackage{subfigure}
\usepackage{verbatim}
\usepackage{amsthm} 
\usepackage{tabularx}
\usepackage{enumitem}    
\usepackage{array}
\usepackage{orcidlink}

\begin{document}

\title{Hamiltonian approach to isospectrality in nonrotating black holes}

\author{Guillermo A. Mena Marug\'an\,\orcidlink{0000-0003-3378-9610}
}
\email{mena@iem.cfmac.csic.es}
\affiliation{Instituto de Estructura de la Materia, IEM-CSIC, C/ Serrano 121, 28006 Madrid, Spain}

\author{Andr\'es M\'inguez-S\'anchez\,\orcidlink{0000-0003-3103-3182}
}
\email{andres.minguez.sanchez@csic.es}
\altaffiliation{Affiliated to the PhD Program, Departamento de F\'isica Te\'orica, Universidad Complutense de Madrid, 28040 Madrid, Spain}
\affiliation{Instituto de Estructura de la Materia, IEM-CSIC, C/ Serrano 121, 28006 Madrid, Spain}

\author{Carlos F. Sopuerta\,\orcidlink{0000-0002-1779-4447}
}
\email{carlos.f.sopuerta@csic.es}
\affiliation{Institut de Ci\`encies de l'Espai (ICE, CSIC), Campus UAB, Carrer de Can Magrans s/n, 08193 Cerdanyola del Vall\`es, Spain}
\affiliation{Institut d'Estudis Espacials de Catalunya (IEEC), Edifici Nexus, Carrer del Gran Capit\`a 2-4, despatx 201, 08034 Barcelona, Spain}

\begin{abstract}
In this work, we revisit the isospectrality of black hole perturbations from a Hamiltonian perspective. For a nonrotating, uncharged system, we present a convenient canonical description of the background degrees of freedom that facilitates comparison with the usual Lagrangian formulation. This allows us to identify phase-space counterparts of the superpotential and the algebraically special modes and show that they satisfy the expected isospectral relations, at least on shell or off shell modulo some subtle lapse redefinitions. Finally, we investigate the extent to which these results persist in theories that effectively modify general relativity. In this setting, we formally quantify the corresponding deviations from the classical relations and the associated potential violations of isospectrality to first order in the modifications.
\end{abstract}

\maketitle

\section{Introduction\label{S-I}}

Isospectrality refers to the property whereby distinct dynamical systems can possess identical spectra despite being described by different fundamental equations \cite{GT}. In black hole (BH) perturbation theory, this phenomenon is particularly remarkable, as gravitational perturbations belonging to different geometrical sectors and governed by apparently unrelated differential equations can nevertheless exhibit the same spectrum of quasinormal modes \cite{N,Chandra2}. Although isospectrality is most commonly discussed in terms of the quasinormal mode spectrum, corresponding to the resonant part of the spectrum of a non-Hermitian Hamiltonian, the equivalence also extends to the continuous spectrum. In particular, the scattering properties of the two sectors coincide, yielding identical reflection and transmission probabilities and, consequently, identical greybody factors. For the Schwarzschild BH, this equivalence was first established by Chandrasekhar \cite{Chandra3,MTBH}, who showed that the axial and polar perturbations, described respectively by the Regge-Wheeler \cite{RW} and Zerilli \cite{Z} master equations, were isospectral. Following this discovery, related spectral correspondences were explored in more general settings, including the Reissner–Nordström and Kerr spacetimes, where the analysis became considerably more involved owing to the presence of charge \cite{RN} and rotation \cite{T}. At present, the study of isospectrality remains an active area of research, with ongoing efforts to understand its origin and its persistence in more general gravitational settings \cite{EGHS,CKMS}.

Interest in BH perturbations has grown substantially in recent years, driven in part by the advent of gravitational-wave astronomy and the prospect of increasingly precise observations of BH mergers \cite{E1,LISA,E2,ET,Bea,Aea}. This empirical progress has been accompanied by significant advances in the theoretical description, including the treatment of rotating BHs \cite{KS,LSC}, higher-order perturbative effects \cite{NI,SPM}, and gravitational self-force \cite{PW,BP}. In parallel, the study of quasinormal modes has revealed a rich mathematical structure underlying the perturbative equations. In the Schwarzschild case, for instance, the relation between the axial and polar master functions can also be understood in terms of Darboux transformations, superpotentials, or algebraically special modes \cite{D,GJK,YY,LS1,LS2}. These developments have also prompted the study of whether such structures survive in theories beyond general relativity, where modifications of Einsteinian gravity can lead to a breaking of isospectrality \cite{STGYS,GMT}. 

A complementary perspective on the field is provided by Hamiltonian methods \cite{ADM}. Early attempts in this direction established a canonical framework for the gravitational degrees of freedom \cite{BMM1,BMM2,B,BM}. However, their proposed formulation treated both time and the radial coordinate as separate but relevant evolution directions, making it difficult to construct a unified phase-space description of the interior and exterior regions. More recent developments have overcome this issue by introducing a single evolution direction that is timelike inside the BH and radial outside \cite{MM1,MM2}\footnote{For other later works on BH backgrounds that identify the interior timelike direction as the dynamical direction of evolution, see e.g. Refs. \cite{Eter,Uncertain}}. This construction offers a more consistent canonical framework across both regions, and describes the axial \cite{MM1,LMMS1} and polar \cite{MM2,LMMS2,LMM} gravitational degrees of freedom to quadratic perturbative order in the action, and therefore to linear order in the associated field equations. Importantly, it also permits the background geometry to be treated dynamically, while the perturbations can be analyzed off-shell and without imposing any gauge condition on them \cite{MM1,MM2,L,CMM}. For the notion of gauge invariance for perturbations in general relativity, see e.g. Refs. \cite{Bardeen,Sasaki}.

Within this emerging formulation, an appropriate choice of canonical variables isolates the physically relevant degrees of freedom into a perturbative gauge-invariant sector in which the configuration variables are, up to normalization by constant factors, the Cunningham-Price-Moncrief (CPM) \cite{CPM} and Zerilli-Moncrief (ZM) \cite{Z,M} master functions. Their conjugate momenta are chosen to make sure that the perturbative dynamics is governed by a quadratic Hamiltonian in the perturbations with a diagonal structure reminiscent of a collection of harmonic oscillators \cite{LMMS1,LMMS2}. The effective mass terms of these oscillators are directly related to generalized forms of the Regge-Wheeler and Zerilli potentials \cite{LMM}. Consequently, the Hamilton equations for each gauge-invariant canonical pair combine to reproduce generalized forms of the axial and polar master equations. This formulation therefore provides a direct phase-space realization of the usual perturbative description of the dynamics of nonrotating, uncharged BHs.

Despite being one of the most remarkable features of BH perturbations, isospectrality is known to be a fragile property, depending sensitively on the structure of the perturbation equations. Given the extensive literature on the subject, we cite only a representative selection of works covering different approaches and physical contexts \cite{STGYS,JPMAS1,JPMAS2,CKMBMM,CD,BS}. Understanding the origin of isospectrality and the conditions under which it holds or breaks down requires a framework that makes the relations between the axial and polar sectors explicit. In this regard, a Hamiltonian formulation provides a natural setting, as it allows these relations to be characterized directly in phase space. Remarkably, recent work has shown that the well-known Darboux transformations \cite{D} underlying the isospectral relations admit a natural interpretation in terms of a distinguished class of canonical transformations that preserve the diagonal structure of the Hamiltonian \cite{LMMS1,LMMS2}. In the standard Lagrangian formulation, these isospectral relations can alternatively be encoded through a superpotential connecting the two potentials or, equivalently, through algebraically special solutions of the corresponding master equations at distinguished quasinormal frequencies \cite{GJK,YY}. These two descriptions are intimately related by the factorization of the second-order master equations in terms of a first-order Riccati equation, and vice versa, and can therefore be viewed as different manifestations of the same fundamental structure. Building on this Darboux-based perspective, our first objective is to determine how these constructions are realized in phase space. We then extend the analysis by exploiting the dynamical treatment of the background to investigate how these structures change when the gravitational dynamics is deformed beyond general relativity. Rather than computing the quasinormal-mode spectra explicitly, as has been extensively done in the literature, we focus on the fundamental relations between the corresponding master equations. Since the spectrum depends not only on the differential equations but also on the boundary conditions imposed on them, the existence of such relations does not by itself guarantee spectral equivalence for an arbitrary boundary-value problem. Nevertheless, their breakdown provides a direct diagnostic of the loss of the mathematical structure supporting isospectrality.

To address these questions in a step-by-step manner, we organize the manuscript as follows. In Sec. \ref{S-II}, we briefly review the perturbative description of BHs and compare the Lagrangian and Hamiltonian approaches, introducing the notation and basic equations used throughout the paper. In Sec. \ref{S-III}, we develop an alternative Hamiltonian description of the background geometry and derive a new constant of motion associated with its asymptotic normalization. Additionally, we reformulate the perturbative Hamiltonian description in order to make the connection with the Lagrangian formulation more transparent. We then use this result in Sec. \ref{S-IV} to identify a phase-space combination from which the axial and polar potentials can be obtained and show that it reproduces the superpotential of the Lagrangian description, up to terms proportional to the background Hamiltonian constraint. We further extend this phase-space construction to determine canonical counterparts of the algebraically special modes. In Sec. \ref{S-V}, we exploit the generality of the Hamiltonian framework to investigate how the canonical relations supporting isospectrality are modified when the background dynamics includes effective corrections that depart from general relativity. Without imposing overly restrictive assumptions, we determine the conditions required for the isospectral relations to persist and characterize their deviations at first order in the modifications. Finally, our conclusions and prospects for further work are presented in Sec. \ref{S-VI}. Throughout the paper, we work in units in which the speed of light and Newton's gravitational constant are set to one.

\section{Perturbative gauge invariants for nonrotating black holes\label{S-II}}

We begin by introducing the basic framework for describing BH perturbations. In the standard formulation, physical perturbations are encoded in gauge-invariant master functions that satisfy second-order differential equations known as master equations \cite{RW,Z}. The solutions to these master equations encode the full physical content of the perturbations, in the sense that all gravitational-wave observables can be derived from them. We first review perturbations of the Schwarzschild spacetime within the Lagrangian framework and then reformulate the description in Hamiltonian terms to establish the canonical formalism employed in the subsequent analysis.

\subsection{Lagrangian formulation}

Linear perturbations of the Schwarzschild spacetime were initially formulated within the Lagrangian framework. This formulation established that the physical perturbative degrees of freedom can be fully characterized by two gauge-invariant quantities, commonly referred to as master functions \cite{CPM,M}. Their dynamics, derived from the linearized Einstein equations, is governed by the following master equations:
\begin{equation}
    \label{S2-eq1}
    \big[\partial^2_{\chi^*} + \omega^2 - V_{\text{RW}}(\chi)\big]\psi^{\ell,m}_{\text{CPM}}(\omega,\chi) = 0, \qquad \big[\partial^2_{\chi^*} + \omega^2 - V_{\text{Z}}(\chi)\big]\psi^{\ell,m}_{\text{ZM}}(\omega,\chi) = 0,
\end{equation}
where $\chi^*$ is the tortoise counterpart of the variation parameter $\chi$ of the master functions (see Eq. \eqref{S2-eq3} below for its explicit expression). In the construction of these formulas, the background spherical symmetry is exploited to separate the angular dependence of the perturbations through an expansion in spherical harmonics, with $\ell$ and $m$ denoting the angular and azimuthal mode numbers, respectively, where $\ell \geq 2$ (modes with $\ell < 2$ contain pure gauge contributions and therefore are not included\footnote{See Refs. \cite{MM1,MM2} for a detailed explanation.}) and $-\ell \leq m \leq \ell$. The distinct parity properties of the harmonics naturally decompose the perturbations into axial and polar sectors \cite{D}. The harmonic modes are then further decomposed (with respect to a coordinate direction orthogonal to $\chi$ in the metric induced on the bidimensional set of spherically symmetric orbits) by means of a Fourier expansion, with $\omega$ denoting the Fourier frequency. After forming suitable combinations of the Fourier modes, the physical degrees of freedom in the axial sector can be encoded in the CPM master function, while their polar counterparts are encoded in the ZM master function. This procedure results in two sets of second-order differential equations for the master functions, expressed in terms of the coordinate $\chi$ (which serves as a radial coordinate in the exterior of the BH, and as a time in the interior), or equivalently in terms of the associated tortoise coordinate $\chi^*$, with potentials given by the Regge-Wheeler and Zerilli expressions \cite{RW,Z}, respectively
\begin{equation}
    \label{S2-eq2}
    \begin{aligned}
        &V_{\text{RW}}(\chi) = \frac{f(\chi)}{\chi^2}\left[\ell(\ell+1) - \frac{6M}{\chi}\right],\\
        &V_{\text{Z}}(\chi) = \frac{f(\chi)}{\chi^2}\frac{1}{\Lambda_{\ell}(\chi)^2}\left[(\ell+2)^2(\ell-1)^2\left(\ell(\ell+1) + \frac{6M}{\chi}\right) + \left(\frac{6M}{\chi}\right)^2\left((\ell+2)(\ell-1) + \frac{2M}{\chi}\right)\right],
    \end{aligned}
\end{equation}
where $M$ denotes the BH mass, and the remaining functions are defined as
\begin{equation}
    \label{S2-eq3}
    \chi^*(\chi) = \chi + 2M\ln\left|\frac{\chi}{2M} - 1\right|, \qquad f(\chi) = 1 - \frac{2M}{\chi}, \qquad \Lambda_{\ell}(\chi) = (\ell+2)(\ell-1) + \frac{6M}{\chi}.
\end{equation}
Together, these expressions constitute a consistent perturbative formulation of the Schwarzschild BH, providing a framework in which isospectrality and other fundamental properties can be studied. Beyond the original works, modern reviews can be found in Refs. \cite{LS2,BCS}.

\subsection{Hamiltonian formulation}

The Hamiltonian formalism of BH perturbations was developed considerably later than the Lagrangian one. Nevertheless, efforts to construct such a canonical description led to a framework in which the background and perturbations are combined into a single phase space, allowing for an off-shell, coordinate-independent description with a dynamical background \cite{MM1,MM2}. Moreover, the generality of this framework permits a consistent formulation in both the interior and exterior regions of the BH, where the Hamiltonian generates time evolution in the former and radial evolution in the latter \cite{MM1,MM2,LMMS1,LMM,LMMS2}.

In phase space, the perturbative degrees of freedom are represented by a collection of six canonical pairs, two associated with the axial contributions and four with the polar contributions:
\begin{equation}
    \label{S2-eq4}
    \{\mathcal{Q}_j^{\mathfrak{n}},\mathcal{P}_{j'}^{\mathfrak{n}'}\} = \kappa\delta_{jj'}\delta^{\mathfrak{n}\mathfrak{n}'},
\end{equation}
where $\kappa=16\pi$ is a constant, the label $j$ runs from $1$ to $6$, and the brackets denote the Poisson brackets defining the perturbative part of the phase-space symplectic structure. The multi-index $\mathfrak{n}=(n,\ell,m)$ collectively denotes the Fourier, angular, and azimuthal mode numbers, respectively, with $n\in\mathbb{Z}$, so that the canonical pairs correspond to the Fourier modes of the BH perturbations. By choosing a suitable set of variables, the description can be effectively reduced to a gauge-invariant sector in which the perturbative dynamics is governed by a relatively simple Hamiltonian:
\begin{equation}
    \label{S2-eq5}
    \kappa\tilde{\mathcal{H}}[\tilde{N}] = \sum_{\mathfrak{n}} \frac{\tilde{N}}{2}|p_c|\left[[\mathcal{P}_1^{\mathfrak{n}}]^2 + (\omega_n^2 - \mathcal{V}_{\text{ax}})[\mathcal{Q}_1^{\mathfrak{n}}]^2 + [\mathcal{P}_3^{\mathfrak{n}}]^2 + (\omega_n^2 - \mathcal{V}_{\text{po}})[\mathcal{Q}_3^{\mathfrak{n}}]^2\right],
\end{equation}
where $\omega_n=2\pi n/L_0$ is a discrete Fourier frequency, with $L_0$ being a fiducial value introduced to avoid infrared subtleties associated with infinite volumes. If necessary, the appropriate continuous limit can reached by suitably letting $L_0$ grow unboundedly. This perturbative Hamiltonian represents the leading-order correction to the Hamiltonian constraint of the theory \cite{LMM}, which in the absence of perturbations is a global zero-mode constraint. At our truncation order, this global Hamiltonian constraint consists of a zeroth-order contribution, referred to as the background Hamiltonian, and the commented second-order correction \cite{MM1,MM2,LMMS1,LMM}. The density weight of the lapse function $\tilde{N}$, and consequently that of the above perturbative Hamiltonian, differs from that usually employed in the canonical formulation of general relativity. The relation between the undensitized and densitized cases in the interior (exterior) region is proportional to the volume of the constant-time (constant-radius) sections, but classically this distinction has no significant effect on the framework.

The canonical pairs labeled by $1$ and $3$ appearing in the perturbative Hamiltonian represent the axial and polar gauge invariants of the theory, respectively. In the present formulation, the configuration variables can be identified with well-known master functions, namely, the CPM and ZM invariants\footnote{The exact relations involve a normalization factor depending on the angular momentum number, which are omitted here for simplicity, but are explicitly given in Ref. \cite{LMM}.}, while their conjugate momenta are chosen such that the perturbative Hamiltonian takes an oscillator-like form. The remaining pairs, labeled by $2$ and by $4$ to $6$, constitute the axial and polar gauge sectors, respectively. In particular, each pair contains a configuration variable, associated with a pure gauge contribution, and a conjugate momentum, which is identified with a perturbative constraint (in Abelianized form, see Refs. \cite{LMM,LMMS2}). Physically, these constraints generate perturbative gauge transformations. Since they are part of the set of perturbative canonical variables, the remaining pairs, namely $1$ and $3$, are gauge invariant by construction \cite{CMM}. The potentials
\begin{equation}
    \label{S2-eq6}
    \begin{aligned}
        &\mathcal{V}_{\text{ax}} = \frac{1}{p_c^2}\frac{p_b^2}{L_0^2}\left[\ell(\ell+1) - 6\Omega_b\Omega_c\frac{L_0^2}{p_b^2}\right],\\
        &\mathcal{V}_{\text{po}} = \frac{1}{p_c^2}\frac{p_b^2}{L_0^2}\frac{1}{\Lambda_{\ell}^2}\left[(\ell+2)^2(\ell-1)^2\left(\ell(\ell+1) + 6\Omega_b\Omega_c\frac{L_0^2}{p_b^2}\right) + \left(6\Omega_b\Omega_c\frac{L_0^2}{p_b^2}\right)^2\left((\ell+2)(\ell-1) + 2\Omega_b\Omega_c\frac{L_0^2}{p_b^2}\right)\right],
    \end{aligned}
\end{equation}
where
\begin{equation}
    \label{S2-eq7}
    \Lambda_{\ell} = (\ell+2)(\ell-1) + 6\Omega_b\Omega_c\frac{L_0^2}{p_b^2},
\end{equation}
generalize the standard Regge-Wheeler and Zerilli expressions as functions on the background phase space. More concretely, apart from their dependence on the angular momentum number, as was already the case in Eq. \eqref{S2-eq2}, they are functions of phase-space background quantities, denoted by $\Omega_b$, $\Omega_c$, $p_b$, and $p_c$, the definition of which is given later in this subsection. 

The master equations follow naturally from the perturbative Hamilton equations and, since the background is not specified, take a more general form. By combining the equations for the configuration and momentum variables, the canonical equations can be written as
\begin{equation}
    \label{S2-eq8}
    \big[\partial^2_{\eta} + \omega_n^2 - \mathcal{V}_{\text{ax}}(\chi)\big] \mathcal{Q}_1^{\mathfrak{n}}(\chi) = 0, \qquad \big[\partial^2_{\eta} + \omega_n^2 - \mathcal{V}_{\text{po}}(\chi)\big] \mathcal{Q}_3^{\mathfrak{n}}(\chi) = 0,
\end{equation}
where $\eta$ is a generalized tortoise coordinate related to $\chi$ through the differential relation $\text{d}\eta = \tilde{N}(\chi)|p_c|(\chi)\text{d}\chi$. When computing solutions, the coordinate dependence of the potentials is already encoded in the background trajectories, and the Schwarzschild background provides the particular case in which they reduce to Eq. \eqref{S2-eq1} (in a suitable limit with the discrete Fourier frequencies $\omega_n$ tending to continuous values $\omega$).

In addition to the perturbative degrees of freedom, the phase space contains two canonical pairs describing the background geometry. A convenient choice of variables, which admits a geometrical interpretation, is given by the canonical pairs $(b,p_b)$ and $(c,p_c)$. For the interior geometry, the configuration variables $b$ and $c$ determine the Ashtekar-Barbero $SU(2)$ connection, while their conjugate momenta, $p_b$ and $p_c$, determine its densitized triad \cite{AOS1,AOS2,EGM}.
The exterior geometry can be described through a complexification of the interior canonical variables \cite{AOS2}. In particular, this can be implemented by the complex canonical transformation $(b,p_b)\rightarrow(ib,-ip_b)$, under which the resulting canonical pairs can be naturally associated with the Ashtekar-Barbero $SU(1,1)$ connection and its densitized triad \cite{LMMS1,LMMS2}. Although other approaches are possible for relating the interior and exterior formulations, this complex canonical transformation provides a simple way to move between the geometries of the two BH regions --up to an overall sign change of the evolution parameter to respect certain conventions--, while preserving the same symplectic structure in phase space. For the analysis presented in this manuscript, it is more convenient to work with the exterior variables, for which the Hamiltonian evolution is radial and is governed by
\begin{equation}
    \label{S2-eq9}
    \{b,p_b\} = \gamma, \qquad \{c,p_c\} = 2\gamma, \qquad \tilde{H}_{\text{KS}} [\tilde{N}]= - \frac{L_0}{2} \tilde{N} \left(\Omega_b^2 + 2\Omega_b\Omega_c - \frac{p_b^2}{L_0^2}\right), \qquad \Omega_a = \frac{ap_a}{\gamma L_0} \;\; \text{for} \;\; a = b\; \text{or} \; c,
\end{equation}
where $\gamma$ is the Immirzi parameter \cite{Immirzi}, which appears as a consequence of our choice of canonical pairs as triad-connection variables \cite{AL,Thiemann}. The combinations $\Omega_b$ and $\Omega_c$, known as dilation functions, together with the triad variables, provide a convenient characterization of phase-space functions \cite{AOS1,AOS2}. Using the triad variables, the line element for the exterior geometry can be written as
\begin{equation}
    \label{S2-eq10}
    \text{d}s^2 = \frac{p_b^2(\chi)}{L_0^2}\left[-\frac{1}{|p_c|(\chi)}\text{d}\zeta^2 + |p_c|(\chi)\tilde{N}(\chi)^2\text{d}\chi^2\right] + |p_c|(\chi)(\text{d}\theta^2 + \sin^2\theta\text{d}\phi^2).
\end{equation}
The extrinsic curvature, which is conjugate to the metric, can be expressed similarly in terms of the Ashtekar-Barbero connection. Thus, the canonical description of the BH background can also be formulated using the more conventional metric-based variables. Nevertheless, the resulting formulation is independent of whether connection-based or metric-based variables are employed. To avoid confusing the time and radial coordinates when switching between the two BH regions, we have used a nonstandard notation for them (different, e.g., from $t$ and $r$). 

A particularly interesting solution to the background equations for classical general relativity is
\begin{equation}
    \label{S2-eq11}
    \Omega_c(\chi) = M, \qquad |p_c|(\chi) =\chi^2, \qquad \Omega_b(\chi) =\chi f(\chi) , \qquad \frac{p_b^2(\chi)}{L_0^2} = \chi^2f(\chi).
\end{equation}
It can be obtained when the lapse is fixed to $\tilde{N}_{\text{Sch}} = L_0^2/p_b^2$ and leads to trajectories that reproduce the Schwarzschild solution. In particular, substituting the background solution \eqref{S2-eq11} into Eq. \eqref{S2-eq10} yields the Schwarzschild line element.

Together, these background \cite{AOS1,AOS2,AO,EGM} and perturbative \cite{MM1,MM2} descriptions constitute the canonical formulation for the perturbative BH geometry (at leading nontrivial order). Both sectors are treated dynamically and are related through the Hamiltonian constraint. As in the Lagrangian framework, the perturbations can be described in terms of master functions and master equations \cite{LMM}, but with the added benefit of the greater generality of the canonical formulation \cite{LMMS2}.

\section{Alternative background formulation\label{S-III}}

Before delving further into the isospectral analysis of quasinormal modes, it is helpful to introduce an alternative description of the background variables that clarifies the connection between the Hamiltonian and Lagrangian approaches. This section is devoted to reorganizing the original set of phase-space variables, comprising the dilation functions and the triads, into a simpler set that admits a more intuitive interpretation.

In general, identifying additional constants of motion is expected to simplify the parametrization of the phase-space dynamics. The initial background description revealed two such constants, both of which are readily identified. The first is the background Hamiltonian itself, while the second is $\Omega_c$, for which the physical interpretation is ultimately related to the BH mass, as shown in Eq. \eqref{S2-eq11}. In their search for further constants of motion, previous studies identified a quantity for which the Poisson bracket with the background Hamiltonian vanishes on shell \cite{LMMS1}. Although this quantity was not explicitly named in the original work, we denote it here by
\begin{equation}
    \label{S3-eq12}
    \Pi^2 = \frac{1}{|p_c|}\left(\frac{1}{\Omega_b}\frac{p_b^2}{L_0^2}\right)^2, \qquad \{\Pi^2,\tilde{H}_{\text{KS}}[\tilde{N}]\} = -\frac{4}{L_0}\frac{\Pi^2}{\Omega_b}\tilde{H}_{\text{KS}}[\tilde{N}].
\end{equation}
Other constants of motion have also been identified in the literature \cite{AOS2,EGM}, including
\begin{equation}
    \label{S3-eq13}
    O_b = -\frac{1}{2}\left(\Omega_b - \frac{1}{\Omega_b}\frac{p_b^2}{L_0^2}\right),  \qquad \{O_b,\tilde{H}_{\text{KS}}[\tilde{N}]\} = -\frac{1}{\Omega_b^2}\frac{p_b^2}{L_0^2}\tilde{H}_{\text{KS}}[\tilde{N}].
\end{equation}
On shell, this last constant also coincides with the BH mass. However, a closer examination reveals that $O_b$ is not functionally independent of the first two constants of motion, as it can be expressed as
\begin{equation}
    \label{S3-eq14}
    O_b = \Omega_c + \frac{1}{\Omega_b}\frac{\tilde{H}_{\text{KS}}}{L_0}.
\end{equation}
It does not represent a new independent constant, but rather a combination of the two quantities already identified. By contrast, $\Pi^2$ is indeed a functionally independent expression. Although this observation does not constitute a complete proof, it was already suggested by the fact that $\Pi^2$ is the only one of the three quantities that depends explicitly on $p_c$. Its physical interpretation is somewhat more subtle than that of $\Omega_c$, since it takes a value equal to one on solutions. To better understand this result, we can rewrite $\Pi^2$ in terms of the metric functions ($g_{\mu\nu}$, with $\mu$ and $\nu$ being spacetime indices) as
\begin{equation}
    \label{S3-eq15}
    \Pi^2\big|_{\text{Sch}}(\chi) = g^{\text{Sch}}_{\zeta\zeta}(\chi)g^{\text{Sch}}_{\chi\chi}(\chi)\left[\partial_{\chi}\sqrt{g^{\text{Sch}}_{\theta\theta}(\chi)}\right]^{-2},
\end{equation}
where the labels ${\text{Sch}}$ stand for evaluation on the Schwarzschild solution. For this solution, the last factor is identically equal to one, whereas the first two terms, which correspond to the product of the nonangular metric components, reduce to a normalization value fixed to unity by the requirement of asymptotic flatness. Consequently, the third constant of motion is directly related to the asymptotic normalization of the metric.

Given these three constants of motion, any function on phase space can be expressed in terms of them and a single additional phase-space quantity. To identify its expression, we seek a function that reduces on shell to the radial coordinate in the Schwarzschild exterior. To this end, we first introduce an auxiliary object that reproduces, off shell, the derivative with respect to this coordinate,
\begin{equation}
    \label{S3-eq16}
    \partial_{\chi} G = \frac{L_0^2}{p_b^2}\frac{1}{\tilde{N}}\{G,\tilde{H}_{\text{KS}}[\tilde{N}]\},
\end{equation}
for any phase-space function $G$. Although the notation may suggest an on-shell construction, we retain it for simplicity, as the quantity itself is well defined off shell. The desired phase-space variable can then be obtained from
\begin{equation}
    \label{S3-eq17}
    \text{X} = \Omega_b + 2\Omega_c, \qquad \partial_{\chi}\text{X} = 1, \qquad \text{X}|_{\text{Sch}}(\chi) = \chi.
\end{equation}
However, the quantity defined in Eq. \eqref{S3-eq12} has a minor drawback. Unlike $\Omega_c$, the Poisson bracket of $\Pi^2$ with the background Hamiltonian vanishes only on shell, rather than identically. Although this poses no physical problem, an identically vanishing Poisson bracket is more convenient. This issue can be resolved by rescaling the conserved quantity by a suitable phase-space function. To determine this function, we use the tools introduced in Eqs. \eqref{S3-eq16} and \eqref{S3-eq17} to solve a first-order differential equation on phase space by quadratures. This yields
\begin{equation}
    \label{S3-eq18}
    \mathbf{\Pi}^2 = \Pi^2e^{\mathcal{I}}, \qquad \mathcal{I} = \ln \left[\frac{\text{F}(\text{X})}{\tilde{\text{F}}(\text{X})}\right] - \ln \left[1 + \frac{2\tilde{H}_{\text{KS}}}{\text{X}^2\text{F}(\text{X})L_0}\right], \qquad \{\mathbf{\Pi}^2,\tilde{H}_{\text{KS}}[\tilde{N}]\} = 0,
\end{equation}
where
\begin{equation}
    \label{S3-eq19}
    \text{F}(\text{X}) = 1 - \frac{2\Omega_c}{\text{X}}, \qquad \tilde{\text{F}}(\text{X}) = \left(\frac{\text{X} - \Omega_c - \sqrt{\Omega_c^2 - 2[\tilde{H}_{\text{KS}}/L_0]}}{\text{X} - \Omega_c + \sqrt{\Omega_c^2 - 2[\tilde{H}_{\text{KS}}/L_0]}}\right)^{\Omega_c/\sqrt{\Omega_c^2 - 2[\tilde{H}_{\text{KS}}/L_0]}}.
\end{equation}
The phase-space factor $e^{\mathcal{I}}$ ensures that the Poisson bracket of $\mathbf{\Pi}^2$ with the background Hamiltonian vanishes identically. The notation used in Eq. \eqref{S3-eq19} is not coincidental, as the expression on the left mimics the usual Schwarzschild function introduced in Eq. \eqref{S2-eq3}. Furthermore, the function on the right reduces on shell to the expression on the left. Compared with Eq. \eqref{S3-eq12}, the corrected expression for the constant of motion is more complicated, but it maintains the same physical interpretation. 

The triad-connection variables used to construct the Hamiltonian formulation can now be expressed in terms of $\text{X}$ and the three constants of motion, namely, the background Hamiltonian, the asymptotic normalization, and the BH mass, as follows:
\begin{equation}
    \label{S3-eq20}
    |p_c| = \frac{\text{X}^2}{\mathbf{\Pi}^2}\left[\frac{\text{F}(\text{X})}{\tilde{\text{F}}(\text{X})} + \frac{2[\tilde{H}_{\text{KS}}/L_0]}{\text{X}^2\tilde{\text{F}}(\text{X})}\right], \qquad \Omega_b = \text{X}\text{F}(\text{X}), \qquad \frac{p_b^2}{L_0^2} = \text{X}^2\text{F}(\text{X}) + 2\frac{\tilde{H}_{\text{KS}}}{L_0}.
\end{equation}
These relations are closer in form to those of the Schwarzschild solution given in Eq. \eqref{S2-eq11}. Moreover, on shell, they coincide with them, without requiring any gauge choice. This new characterization of phase-space functions is also useful at the perturbative level, as it allows us to express quantities such as the axial and polar potentials introduced in Eq. \eqref{S2-eq6} in a form closer to their standard Lagrangian counterparts. In particular
\begin{equation}
    \label{S3-eq21}
    \begin{aligned}
        &\mathcal{V}_{\text{ax}} = \frac{\mathbf{\Pi}^4\tilde{\text{F}}(\text{X})^2}{\text{X}^2\text{F}(\text{X}) + 2[\tilde{H}_{\text{KS}}/L_0]} \Bigg[\ell(\ell+1) - \frac{6\Omega_c}{\text{X} + 2\tilde{H}_{\text{KS}}/[\text{X}\text{F}(\text{X})L_0]}\Bigg],\\
        &\begin{aligned}
            \mathcal{V}_{\text{po}} &= \frac{\mathbf{\Pi}^4\tilde{\text{F}}(\text{X})^2}{\text{X}^2\text{F}(\text{X}) + 2[\tilde{H}_{\text{KS}}/L_0]}\frac{1}{\Lambda_{\ell}^2} \Bigg[(\ell+2)^2(\ell-1)^2\left(\ell(\ell+1) + \frac{6\Omega_c}{\text{X} + 2\tilde{H}_{\text{KS}}/[\text{X}\text{F}(\text{X})L_0]}\right)\\ 
            &+ \left(\frac{6\Omega_c}{\text{X} + 2\tilde{H}_{\text{KS}}/[\text{X}\text{F}(\text{X})L_0]}\right)^2\left((\ell+2)(\ell-1) + \frac{2\Omega_c}{\text{X} + 2\tilde{H}_{\text{KS}}/[\text{X}\text{F}(\text{X})L_0]}\right)\Bigg],
    \end{aligned}
    \end{aligned}
\end{equation}
where now
\begin{equation}
    \label{S3-eq22}
    \Lambda_{\ell} = (\ell+2)(\ell-1) + \frac{6\Omega_c}{\text{X} + 2\tilde{H}_{\text{KS}}/[\text{X}\text{F}(\text{X})L_0]}.
\end{equation}
When regarded as functions on phase space, the potentials in Eq. \eqref{S3-eq21}, which correspond to those in Eq.\eqref{S2-eq6} but expressed in terms of different phase-space functions, provide a clearer picture of their more general structure compared with their Lagrangian counterparts. Nevertheless, both formulations reduce to the same functional form when the background satisfies the on-shell condition.

\section{Superpotential and algebraically special modes\label{S-IV}}

Having discussed the derivation of the master equations, we now turn to their analysis. Equations \eqref{S2-eq1} and \eqref{S2-eq7} have a similar structure, but differ in one important respect: solving the latter requires fixing the lapse, which, in the exterior (interior) geometry, amounts to choosing a specific radial (timelike) coordinate. Once this choice has been made, the equations can be solved for each angular momentum number, leading to the well-known quasinormal-mode problem \cite{N}. Quasinormal modes describe damped gravitational perturbations and are characterized by a discrete spectrum of complex frequencies determined by the relevant boundary conditions \cite{N}. Remarkably, despite the different functional forms of the Regge–Wheeler and Zerilli potentials, the corresponding quasinormal-mode spectra coincide. This property, known as isospectrality, was elucidated by Chandrasekhar, who demonstrated that the axial and polar sectors are closely related despite their apparently different formulations \cite{Chandra3,MTBH}.

For Schwarzschild spacetime, quasinormal modes have been extensively studied within the Lagrangian formulation, whereas their Hamiltonian counterpart remains largely unexplored. This motivates the question of how much information about the perturbative spectrum can be extracted directly from the Hamiltonian framework, without fixing the lapse or solving the background dynamics. To address this question, we focus on two closely related concepts that illuminate the connection between the axial and polar sectors: the superpotential, which relates the Regge–Wheeler and Zerilli potentials through a Riccati equation, and algebraically special modes, particular solutions which provide a complementary way of characterizing this relation. Neither concept, on its own, guarantees isospectrality; rather, both reveal important aspects of the mathematical structure underlying the relationship between the axial and polar sectors. We begin by reviewing these concepts in the Lagrangian formulation, which will serve as a reference for developing their canonical counterparts.

\subsection{Lagrangian formulation}

For simplicity, we first adopt the Lagrangian approach and introduce the following functions:
\begin{equation}
    \label{S4-eq23}
    W(\chi) = -\frac{1}{12M}\frac{(\ell+2)!}{(\ell-2)!} - \frac{6M}{\chi^2}\frac{f(\chi)}{\Lambda_{\ell}(\chi)}, \qquad \beta = -\left[\frac{1}{12M}\frac{(\ell+2)!}{(\ell-2)!}\right]^2.
\end{equation}
In terms of these quantities, the Regge–Wheeler and Zerilli potentials can be expressed as
\begin{equation}
    \label{S4-eq24}
    V_{\text{RW}}(\chi) = W(\chi)^2 + \partial_{\chi^*}W(\chi) + \beta, \qquad V_{\text{Z}}(\chi) = W(\chi)^2 - \partial_{\chi^*}W(\chi) + \beta.
\end{equation}
The function $W(\chi)$, known as the superpotential, provides a direct relation between the two potentials. It also defines a transformation that maps solutions of the axial master equation to solutions of the polar master equation and vice versa \cite{BCS}. In terms of the CPM and ZM invariants, this transformation can be written as
\begin{equation}
    \label{S4-eq25}
    \psi^{\ell,m}_{\text{CPM}}(\omega,\chi) = -\frac{1}{\sqrt{\omega^2 - \beta}}[\partial_{\chi^*} + W(\chi)]\psi^{\ell,m}_{\text{Z}}(\omega,\chi) , \qquad \psi^{\ell,m}_{\text{Z}}(\omega,\chi) = \frac{1}{\sqrt{\omega^2 - \beta}}[\partial_{\chi^*} - W(\chi)]\psi^{\ell,m}_{\text{CPM}}(\omega,\chi).
\end{equation}
With the chosen normalization, these relations define mutually inverse Darboux transformations, provided that the denominators are nonvanishing and that the master functions satisfy their respective master equations. However, the existence of such transformations does not, by itself, guarantee isospectrality. For the two sectors to share the same spectrum, the transformation must be compatible with the boundary conditions defining the spectral problem and must remain well defined at the relevant frequencies \cite{GV}. In particular, the algebraically special frequencies, for which the denominators vanish, require separate consideration because the transformations become singular.

To understand the connection between the superpotential and these special frequencies, we now consider algebraically special modes. Their relevance stems from the fact that particular solutions of the master equations can be used as seed solutions to construct the superpotential and, consequently, the Darboux transformation in Eq. \eqref{S4-eq25}. For simplicity, we focus on the axial sector. Given a nonvanishing solution of the Regge–Wheeler equation at a fixed frequency, we can formally define
\begin{equation}
    \label{S4-eq26}
    u(\chi) = e^{\int^{\chi^*(\chi)}\,W(x)\text{d}x}, \qquad W(\chi) = \partial_{\chi^*}[\ln u(\chi)].
\end{equation}
Substituting the right-hand side expression into the axial master equation yields a Riccati equation for the logarithmic derivative of the seed solution. Conversely, a solution of the Riccati equation determines a solution of the axial master equation through exponentiation and simultaneously defines a Darboux transformation. At this stage, the construction is not restricted to algebraically special modes, and the resulting transformation relates the Regge–Wheeler potential to a partner potential that need not coincide with the Zerilli potential. To recover the latter, the Riccati equation must reproduce the specific factorization of the Regge–Wheeler potential given in Eq. \eqref{S4-eq24}. This requirement fixes the seed frequency through $w_*^2=\beta$, thereby selecting the algebraically special solution, with explicit expression and associated frequency given by
\begin{equation}
    \label{S4-eq27}
    u(w_{*},\chi) = \frac{1}{2}\Lambda_{\ell}(\chi)e^{-iw_{*}\chi^*(\chi)}, \qquad w_{*} = -\frac{i}{12M}\frac{(\ell+2)!}{(\ell-2)!}.
\end{equation}

Among the complex quasinormal-mode frequencies, the algebraically special frequencies are distinguished by being purely imaginary. Taking the logarithmic derivative of this solution yields the superpotential defined in Eq. \eqref{S4-eq23}, establishing a direct link between the algebraically special modes and the Darboux transformations relating the two sectors. The corresponding master equation and Riccati equation then take the form
\begin{equation}
    \label{S4-eq28}
    \big[\partial^2_{\chi^*} + w^2_{*} - V_{\text{RW}}(\chi)\big]u(w_{*},\chi) = 0, \qquad \partial_{\chi^*}W(\chi) + W(\chi)^2 + w^2_{*} - V_{\text{RW}}(\chi) = 0.
\end{equation}
The polar master equation also admits algebraically special solutions \cite{GJK}. Applying the same procedure to the polar sector yields
\begin{equation}
    \label{S4-eq29}
    v(w_{*},\chi) = \frac{1}{u(w_{*},\chi)}, \qquad \big[\partial^2_{\chi^*} + w^2_{*} - V_{\text{Z}}(\chi)\big]v(w_{*},\chi) = 0.
\end{equation}
Taken together, these results establish the connection between the superpotential, Darboux transformations, and algebraically special modes within the Lagrangian formulation. In particular, they show how the superpotential can be reconstructed from a distinguished solution of the axial or polar master equation and provide a natural starting point for the Hamiltonian analysis. A more detailed discussion of these topics can be found in Refs. \cite{GJK,YY,LS1,LS2}.

\subsection{Hamiltonian formulation}

Following the same motivation as in the Lagrangian description, we introduce the following phase-space expressions:
\begin{equation}
    \label{S4-eq30}
    \mathcal{W} = -\frac{\Pi^2}{12\Omega_c}\frac{(\ell+2)!}{(\ell-2)!} - \frac{1}{|p_c|}\frac{L_0^2}{p_b^2}\frac{1}{\Lambda_{\ell}}6\Omega_b^2\Omega_c, \qquad \mathfrak{b} = -\left[\frac{\Pi^2}{12\Omega_c}\frac{(\ell+2)!}{(\ell-2)!}\right]^2.
\end{equation}
In terms of these quantities, we can construct the Hamiltonian counterparts of the Regge–Wheeler and Zerilli potentials introduced in Eq. \eqref{S2-eq6}. These potentials are expected to satisfy relations analogous to those in Eqs. \eqref{S4-eq24}, now formulated as combinations between phase-space functions that hold off shell. Explicitly,
\begin{equation}
    \label{S4-eq31}
    \begin{aligned}
        &\mathcal{V}_{\text{ax}} = \mathcal{W}^2 + \partial_{\eta}\mathcal{W} + \mathfrak{b} - \frac{1}{|p_c|^2}\frac{1}{\Lambda_{\ell}^2}8\Omega_b\Omega_c\frac{L_0^2}{p_b^2}\left(2(\ell+2)(\ell-1) + 9\Omega_b\Omega_c\frac{L_0^2}{p_b^2}\right)\frac{4\tilde{H}_{\text{KS}}}{L_0} + \frac{\Pi^2}{3\Omega_b\Omega_c}\frac{(\ell+2)!}{(\ell-2)!}\frac{\tilde{H}_{\text{KS}}}{L_0},\\
        &\mathcal{V}_{\text{po}} = \mathcal{W}^2 - \partial_{\eta}\mathcal{W} + \mathfrak{b} + \frac{1}{|p_c|^2}\frac{1}{\Lambda_{\ell}^2}8\Omega_b\Omega_c\frac{L_0^2}{p_b^2}\left(2(\ell+2)(\ell-1) + 3\Omega_b\Omega_c\frac{L_0^2}{p_b^2}\right)\frac{4\tilde{H}_{\text{KS}}}{L_0} - \frac{\Pi^2}{3\Omega_b\Omega_c}\frac{(\ell+2)!}{(\ell-2)!}\frac{\tilde{H}_{\text{KS}}}{L_0}.
    \end{aligned}
\end{equation}
To re-express these relations, we also introduce a phase-space derivative, analogous to the one defined in Eq. \eqref{S3-eq16}, designed to reproduce, off shell, the derivative with respect to the generalized tortoise coordinate. For an arbitrary phase-space function $G$, it is defined as
\begin{equation}
    \label{S4-eq32}
    \partial_{\eta} G = \mathbf{\Pi}^2\tilde{\text{F}}(\text{X}) \partial_{\chi} G.
\end{equation}
However, a direct inspection of Eq. \eqref{S4-eq31} shows that these relations are not exactly like Eq. \eqref{S4-eq24} off shell. More specifically, additional terms proportional to the background Hamiltonian prevent the potentials from satisfying the expected kind of relations identically throughout the entire background phase space, left unconstrained. Nevertheless, as in the examples mentioned earlier, the relations become exact on shell, even without fixing the gauge. One might therefore expect that this discrepancy could be resolved by modifying one of the functions introduced in Eq. \eqref{S4-eq32}, following the same strategy used to adjust the third constant of motion in Eq. \eqref{S3-eq18}. The underlying issue, however, is more subtle. As defined in Eq. \eqref{S2-eq6}, or equivalently in Eq. \eqref{S3-eq21}, the two potentials cease to belong to exactly the same Darboux-related family when considered off shell. This observation does not imply any breakdown of isospectrality. Indeed, as discussed above, the boundary conditions play a crucial role in establishing the correspondence between the axial and polar sectors, and their implementation necessarily requires an on-shell treatment. Consequently, the apparent off-shell mismatch does not obstruct the expected correspondence between the two sectors.

The origin of the discrepancy that prevents the discussed relations from holding exactly off shell can be traced, at least in part, to the original formulation of the potentials \cite{LMM}. Their construction was primarily motivated by the desire to obtain simple expressions in terms of the dilation and triad variables, thereby facilitating the study of quasinormal modes for quantum or effective backgrounds, such as those arising in loop quantum cosmology \cite{AP}, a well-established approach to study simple cosmological and BH spacetimes in loop quantum gravity \cite{AL,Thiemann}, framework in which these variables constructed from triads and connections play a particularly natural role \cite{AOS1,AOS2,AO,EMM}. To achieve this simplification, advantage was taken of a standard freedom in the canonical formulation of perturbation theory in general relativity \cite{MM1,MM2,LMM,LMMS2}: terms proportional to the background Hamiltonian in the perturbative contributions to the constraints can be absorbed into redefinitions of the lapse function, which involve quadratic perturbative corrections to it \cite{M,CMM,MM1,MM2,LMMS2}. This change of lapse is irrelevant if one identifies the original and redefined times, and/or if one restricts the effect of the perturbations in the spacetime metric to linear order. In particular, this procedure can be applied to terms of the aforementioned type in the quadratic perturbative correction to the global Hamiltonian constraint, and therefore in the potential. Although these lapse redefinitions leave the dynamics unchanged on shell, they can alter the off-shell relations between perturbative phase-space functions, as illustrated by the present analysis. These observations highlight the sensitivity of the connection between the Regge–Wheeler and Zerilli potentials to the particular off-shell representatives chosen for their Hamiltonian expressions among those allowed by the discussed freedom. A more systematic investigation of these potentials without the redefinitions introduced to accommodate their off-shell expressions for quantum-geometric applications is left for future work. Such an investigation will also consider alternative definitions of the potentials and of the phase-space derivative, and examine how these choices can affect the corresponding off-shell relations.

Despite these subtleties, as in the Lagrangian description, transformations relating the axial and polar sectors can still be constructed. In the Hamiltonian framework, however, obtaining the analogue of the Darboux transformation is more delicate: rather than taking the form of a differential relation between the master functions, the transformation is realized as a specific linear map on the perturbations that preserves the symplectic structure of phase space while maintaining the diagonal form of the leading perturbative contribution to the Hamiltonian constraint \cite{LMMS1,LMMS2}. We can, for instance, construct the canonical transformation associated with the polar sector, naturally linked to the Zerilli potential, as follows:
\begin{equation}
    \label{S4-eq33}
    \tilde{\mathcal{Q}}_3^{\mathfrak{n}} = - \frac{1}{\sqrt{\omega_n^2 - \mathfrak{b}}}[\mathcal{P}_3^{\mathfrak{n}} + \mathcal{W}\mathcal{Q}_3^{\mathfrak{n}}], \qquad \tilde{\mathcal{P}}_3^{\mathfrak{n}} = -\frac{1}{\sqrt{\omega_n^2 - \mathfrak{b}}}[\mathcal{W}\mathcal{P}_3^{\mathfrak{n}} + (\mathcal{W}^2 + \mathfrak{b} - \omega_n^2)\mathcal{Q}_3^{\mathfrak{n}}].
\end{equation}
The new canonical variables, denoted by a tilde, are subject to the Regge–Wheeler potential, which naturally arises in the axial sector. The coefficients of the transformation are determined by imposing canonicity, requiring the transformed perturbative Hamiltonian to remain diagonal but with the desired Regge-Wheeler potential, and identifying the new momenta with the derivatives of the corresponding configuration variables \cite{LMMS1}. These requirements yield a Hamiltonian of the same form as that in Eq. \eqref{S2-eq5}, but with a common axial potential for both sectors, up to terms proportional to the background Hamiltonian. As noted above, these additional terms can be eliminated through suitable redefinitions. This construction also highlights the close analogy with the Darboux transformations introduced in Eq. \eqref{S4-eq25}. 

In the Lagrangian formalism, algebraically special modes are closely related to superpotentials and Darboux transformations. This suggests that establishing analogous relations in phase space could provide a unified framework for characterizing these modes in the Hamiltonian description. To investigate this possibility in a fully off-shell setting, we first introduce a phase-space function playing the role of the generalized tortoise coordinate. Concretely, we define
\begin{equation}
    \label{S4-eq34}
    \textbf{N} = \frac{1}{\mathbf{\Pi}^2} \int^{\text{X}} \frac{\text{d}\text{Y}}{\tilde{\text{F}}(\text{Y})}, \qquad \partial_{\eta}\textbf{N} = 1, \qquad \textbf{N}|_{\text{Sch}}(\chi) = \chi^*(\chi).
\end{equation}
Unlike the expression for $\text{X}$ in Eq. \eqref{S3-eq17}, the present expression is more involved. Nevertheless, the integral can still be evaluated in terms of Gauss hypergeometric functions, as follows:
\begin{equation}
    \label{S4-eq35}
    \textbf{N} = -\frac{1}{\mathbf{\Pi}^2}\frac{2\Omega_c\text{Z}^{1+\sigma}}{\sigma(1+\sigma)}\,{}_2F_1\left(2,1+\sigma;2+\sigma;\text{Z}\right), \qquad \sigma = \left[1 - \frac{2\tilde{H}_{\text{KS}}}{\Omega_c^2L_0}\right]^{-1/2}, \qquad \text{Z} = \frac{\Omega_c + \sigma(\text{X} - \Omega_c)}{\Omega_c + \sigma(\text{X} + \Omega_c)}.
\end{equation}
Using this definition, together with the expressions in Eqs. \eqref{S2-eq7} and \eqref{S4-eq23}, we can construct phase-space analogues of the algebraically special modes in the axial sector,
\begin{equation}
    \label{S4-eq36}
    \mathfrak{U} = \frac{\Lambda_{\ell}}{2}e^{-i\mathfrak{w}_{*}\textbf{N}}, \qquad \mathfrak{w}_{*} =-\frac{i\Pi^2}{12\Omega_c}\frac{(\ell+2)!}{(\ell-2)!}.
\end{equation}
The algebraically special modes are related to the superpotential in Eq. \eqref{S4-eq30} according to
\begin{equation}
    \label{S4-eq37}
    \mathcal{W} = \partial_{\eta}[\ln\mathfrak{U}] - \frac{4}{|p_c|\Omega_b}\left[i\mathfrak{w}_{*}\textbf{N} + 3\Omega_b\Omega_c\frac{1}{\Lambda_{\ell}}\frac{L_0^2}{p_b^2}\right]\frac{\tilde{H}_{\text{KS}}}{L_0}.
\end{equation}
As expected, the resulting relation does not hold exactly off shell. As in Eq. \eqref{S4-eq31}, additional terms proportional to the background Hamiltonian spoil the exact parallelism with Eq. \eqref{S4-eq26}. Nevertheless, this construction remains useful because it confirms that the expected result is recovered on shell, and also provides a starting point for more detailed investigations. Furthermore, one of the characteristic properties of algebraically special modes can be examined by applying the phase-space analogue of the Hamiltonian master equations to the combination defined in Eq. \eqref{S4-eq36}, replacing the derivatives with the phase-space derivative introduced in Eq. \eqref{S4-eq34}. The resulting relation holds up to terms proportional to the background Hamiltonian. We omit its explicit expression because of its length and its limited relevance to the present discussion. Finally, following the procedure outlined in Eq. \eqref{S4-eq29}, we can also construct phase-space analogues of the algebraically special modes associated with the polar equation by taking the inverse of the axial algebraically special modes. As in the axial case, the corresponding relations do not hold exactly off shell, with deviations proportional to the background Hamiltonian.

\section{Isospectral relations for effective theories\label{S-V}}

In the previous section, we identified how the main structures supporting isospectrality are realized in the Hamiltonian framework of BH perturbation theory. This provides a natural starting point for investigating how these relations are modified when effective corrections to general relativity are introduced\footnote{In this work, we limit our considerations to effective theories derived from general relativity by incorporating modifications that can be treated at the level of an effective Hamiltonian. Our analysis is not directly applicable to other types of effective gravitational theories.}. In particular, such an analysis can help clarify which features of the axial and polar sectors are robust under modifications of the background dynamics and which depend on the specific form of the theory governing them. To this end, we consider effective corrections at the phase-space level, keeping the discussion sufficiently general without committing to any particular effective model. As is commonly assumed in this type of studies of compact objects, we further assume that the dynamics of the perturbations remain approximately close to what is dictated by general relativity.

Within our Hamiltonian formulation, extending the analysis to an effective setting amounts to replacing the background variables by their effective counterparts,
\begin{equation}
    \label{S5-eq38}
    \Omega_c \rightarrow \Omega_c^{\text{eff}}, \qquad p_c \rightarrow p_c^{\text{eff}}, \qquad \Omega_b \rightarrow \Omega_b^{\text{eff}}, \qquad p_b \rightarrow p_b^{\text{eff}}.
\end{equation}
The explicit form of these quantities depends on the model under consideration. Such modifications may also affect the background Hamiltonian and symplectic structure. As a result, the equations of motion are generally modified, and quantities conserved in the original theory may no longer be conserved in the effective one. To parametrize these changes, we assume that the background quantities depend on an external parameter $\epsilon$ controlling the deformation produced by the effective corrections, with the original expressions recovered in the limit where $\epsilon$ goes to zero. For instance, the dilation function of the $c$-sector can be expanded as
\begin{equation}
    \label{S5-eq39}
     \Omega_c^{\text{eff}} = \sum_{n=0}^{\infty} \epsilon^n\Omega_c^{[n]} = \Omega_c + \epsilon \Omega_c^{[1]} + \epsilon^2\Omega_c^{[2]} + \cdots,
\end{equation}
where the case $n=0$ corresponds to the noneffective limit and, for $n>0$, the contributions $\Omega_c^{[n]}$ are phase-space functions of the original variables. Analogous expansions can be introduced for the remaining quantities in Eq. \eqref{S5-eq38}.

The effective description can then be extended from the background dynamics to those of the perturbations. In particular, the master equations are modified by replacing the original evolution with its effective counterpart, and by modifying the potentials to which the perturbations are subject. In principle, there is considerable freedom in defining these effective potentials, provided that they reduce to their original expressions in the limit where $\epsilon$ vanishes. Rather than introducing corrections with complicated functional dependencies, we adopt the simplest prescription, in which the functional form of the original potentials is maintained and the background quantities entering them are merely replaced with their effective counterparts. This prescription, however, does not guarantee that the structures underlying isospectrality are preserved. In particular, the algebraically special modes, the superpotential, and the Darboux transformation cannot, in general, be extended to the effective setting by applying the same substitutions to their original expressions. Consequently, the resulting effective theory does not generically inherit the original isospectral structure, and determining whether any remnant of isospectrality survives requires a dedicated analysis. 

Throughout the following discussion, we formulate the relevant relations as off-shell identities in phase space. For simplicity, we assume that these identities hold exactly at all subdominant orders in the effective corrections, neglecting possible additional terms proportional to the background Hamiltonian except at zeroth order in the effective deformations. To investigate if isospectrality can then be preserved in the effective theory, the first nontrivial question is whether the effective dynamics admits a conserved quantity that can be identified with the constant appearing in the Riccati equation associated with the Darboux transformation. This quantity is directly related to the superpotential constant and, equivalently, to the frequencies of the algebraically special modes. Its existence is therefore essential for constructing the transformation relating the two master equations. For convenience, we assume in what follows that such a conserved quantity exists in the effective description and that its explicit form is known.

A second condition we impose is that the effective potentials constructed according to our prescription satisfy the Darboux relations. Combining the Riccati equation with the difference between the two potentials, we find that, whenever two master equations are related by a Darboux transformation, their potentials must satisfy the following consistency conditions:
\begin{equation}
    \label{S5-eq40}
    \partial_{\eta}^{\text{eff}}\mathcal{W}^{\text{eff}} = \frac{1}{2}\Delta_{-}\mathcal{V}^{\text{eff}}, \qquad \left(\mathcal{W}^{\text{eff}}\right)^2 = \frac{1}{2}\Delta_{+}\mathcal{V}^{\text{eff}} - \mathfrak{b}^{\text{eff}}, \qquad \mathfrak{b}^{\text{eff}} = \frac{1}{2}\Delta_{+}\mathcal{V}^{\text{eff}} - \frac{1}{4}\left(\frac{1}{\Delta_{-}\mathcal{V}^{\text{eff}}}\partial_{\eta}^{\text{eff}}\Delta_{+}\mathcal{V}^{\text{eff}}\right)^2,
\end{equation}
where $\mathcal{W}^{\text{eff}}$ generates the Darboux transformation in the effective theory between the corrected potentials (denoted generically by $\mathcal{V}^{\text{eff}}$), and $\mathfrak{b}^{\text{eff}}$ is its associated constant of motion. The remaining quantities, together with the effective off-shell derivative, are defined as follows: 
\begin{equation}
    \label{S5-eq41}
    \partial_{\eta}^{\text{eff}}G = \frac{1}{|p_c^{\text{eff}}|}\frac{1}{\tilde{N}^{\text{eff}}}\{G,\tilde{H}_{\text{KS}}^{\text{eff}}[\tilde{N}^{\text{eff}}]\}^{\text{eff}}, \qquad \Delta_{+}\mathcal{V}^{\text{eff}} = \mathcal{V}_{\text{ax}}^{\text{eff}} + \mathcal{V}_{\text{po}}^{\text{eff}}, \qquad \Delta_{-}\mathcal{V}^{\text{eff}} = \mathcal{V}_{\text{ax}}^{\text{eff}} - \mathcal{V}_{\text{po}}^{\text{eff}},
\end{equation}
for any phase-space function $G$. The derivative accounts for the effective change in the background variables (and thereafter in the background Hamiltonian) as well as possible modifications of the Poisson-bracket relations. Equation \eqref{S5-eq40} makes clear that the existence of a constant of motion in the effective theory is a necessary condition for establishing the Darboux relations. In the effective description, however, there is no guarantee that the axial and polar potentials obtained with our prescription remain connected by such a relation. The relation will generically fail to hold, even when the required conserved quantity exists. This illustrates the fragility of isospectrality, since relatively small modifications of the dynamics may suffice to destroy the structure responsible for it. Consequently, the effective axial and polar potentials will, in general, belong to different families of Darboux-connected potentials.

This observation suggests an alternative way to characterize the loss of isospectrality: rather than requiring the two effective potentials to satisfy the same Darboux relations, we can quantify their deviation from those relations. For definiteness, we focus on the axial potential, although an analogous analysis can be performed starting from the polar sector. The relevant expressions are
\begin{equation}
    \label{S5-eq42}
    \mathcal{V}_{\text{ax}/\text{po}}^{\text{eff}} = \sum_{n=0}^{\infty} \epsilon^n\mathcal{V}_{\text{ax}/\text{po}}^{[n]}, \qquad \mathcal{W}^{\text{eff}}_{\text{ax}} = \sum_{n=0}^{\infty} \epsilon^n\mathcal{W}^{[n]}_{\text{ax}}, \qquad \mathfrak{b}^{\text{eff}}_{\text{ax}} = \sum_{n=0}^{\infty} \epsilon^n\mathfrak{b}^{[n]}_{\text{ax}},
\end{equation}
where, a priori, the two effective potentials obtained with the aforementioned prescription and the effective axial constant of motion are assumed to be known. As we have noticed, the effective generating function of the axial case need not relate the effective axial potential $\mathcal{V}_{\text{ax}}^{\text{eff}}$ with $\mathcal{V}_{\text{po}}^{\text{eff}}$ as its polar counterpart, but rather to another effective correction of the Zerilli potential, once the effective corrections properly come into the scene. To take into account this fact, we should introduce a different notation for the corresponding effective polar generating function and its associated constant of motion if the polar potential $\mathcal{V}_{\text{po}}^{\text{eff}}$ was to be considered as the starting point. 

The quantities introduced in Eq. \eqref{S5-eq42} satisfy the following differential equation:
\begin{equation}
    \label{S6-eq63}
    \partial_{\eta}^{\text{eff}}\mathcal{W}_{\text{ax}}^{\text{eff}} + \left(\mathcal{W}_{\text{ax}}^{\text{eff}}\right)^2 - \mathcal{V}_{\text{ax}}^{\text{eff}} + \mathfrak{b}_{\text{ax}}^{\text{eff}} = 0,
\end{equation}
possibly up to contributions linear in the background Hamiltonian which have no effective deformations, as we will comment below. Once the above equation is solved, we can qualitatively characterize how $\mathcal{V}_{\text{po}}^{\text{eff}}$ deviates from the combination $\mathcal{V}_{\text{ax}}^{\text{eff}} - 2\partial_{\eta}^{\text{eff}}\mathcal{W}^{\text{eff}}_{\text{ax}}$. Moreover, expanding the resulting expression in powers of $\epsilon$ provides a systematic way to track the development of this deviation order by order. We thus adopt an iterative approach to constructing and analyzing solutions to Eq. \eqref{S6-eq63}. 

To this end, it is convenient to introduce an auxiliary quantity that measures the difference between the effective and original off-shell derivatives, allowing the effective Riccati equation to be rewritten entirely in terms of the original dynamics,
\begin{equation}
    \label{S6-eq64}
    \partial_{\eta}\mathcal{W}_{\text{ax}}^{\text{eff}} + (\partial_{\eta}^{\text{eff}} - \partial_{\eta})\mathcal{W}_{\text{ax}}^{\text{eff}} + \left(\mathcal{W}_{\text{ax}}^{\text{eff}}\right)^2 - \mathcal{V}_{ax}^{\text{eff}} + \mathfrak{b}_{\text{ax}}^{\text{eff}} = 0.
\end{equation}
By construction, the auxiliary difference of derivatives vanishes in the undeformed limit. Its expansion in the deformation parameter $\epsilon$ therefore contains no zeroth-order term and can be formally written as
\begin{equation}
    \label{S6-eq65}
    (\partial_{\eta}^{\text{eff}} - \partial_{\eta})G = \sum_{n=1}^{\infty} \epsilon^nD[G]^{[n]},
\end{equation}
for any phase-space function $G$. Without specifying its explicit construction, we denote by $D[G]^{[n]}$ the contribution at order $n$ in the expansion of the difference between the two derivatives. Using this notation, Eq. \eqref{S6-eq64} can be expanded as follows:
\begin{equation}
    \label{S6-eq66}
    \sum_{n=0}^{\infty}\epsilon^n\partial_{\eta}\mathcal{W}_{\text{ax}}^{[n]} + \sum_{n=1}^{\infty}\sum_{m=0}^{n-1}\epsilon^nD[\mathcal{W}_{\text{ax}}^{[m]}]^{[n-m]} + \sum_{n=0}^{\infty}\sum_{m=0}^{n}\epsilon^n \mathcal{W}_{\text{ax}}^{[m]}\mathcal{W}_{\text{ax}}^{[n-m]} - \sum_{n=0}^{\infty}\epsilon^n\mathcal{V}_{\text{ax}}^{[n]} + \sum_{n=0}^{\infty}\epsilon^n\mathfrak{b}_{\text{ax}}^{[n]} = 0.
\end{equation}

At zeroth order, we recover the original Riccati equation. It is worth noting that, at this zeroth order, we can even include sources in the Ricatti equation \eqref{S6-eq63} proportional to the underformed background Hamiltonian and independent of the deformation parameter $\epsilon$. The generator of the underformed transformation would then satisfy the original Riccati equation up to the contribution of such sources, which vanish on shell. But the deformations of this generator at any orther order, $\mathcal{W}_{\text{ax}}^{[n]}$ with $n>0$, would still have to solve the restriction imposed by Eq. \eqref{S6-eq66}. In fact, at each subsequent order, the resulting restriction is a linear equation that can be solved by an iterative procedure, using the solution obtained at the previous order\footnote{Actually, we could introduce source terms proportional to the underformed background Hamiltonian at all orders, and the iterative procedure would still be applicable. However, we avoid including such higer-order terms for the sake of simplicity, emphasizing that the freedom available in our construction makes them unnecessary in order to formally obtain solutions.}. Explicitly, we get
\begin{equation}
    \label{S6-eq67}
    \partial_{\eta}\mathcal{W}_{\text{ax}}^{[n]} + \sum_{m=0}^{n-1}D[\mathcal{W}_{\text{ax}}^{[m]}]^{[n-m]} + \sum_{m=0}^{n} \mathcal{W}_{\text{ax}}^{[m]}\mathcal{W}_{\text{ax}}^{[n-m]} - \mathcal{V}_{\text{ax}}^{[n]} + \mathfrak{b}_{\text{ax}}^{[n]} = 0, \qquad \text{for} \qquad n>0.
\end{equation}

To illustrate the procedure, we first consider the linear case, for which the differential equation simplifies considerably,
\begin{equation}
    \label{S6-eq68}
    \partial_{\eta}\mathcal{W}_{\text{ax}}^{[1]} + 2\mathcal{W}\mathcal{W}_{\text{ax}}^{[1]} = \mathcal{V}_{\text{ax}}^{[1]} - \mathfrak{b}_{\text{ax}}^{[1]} - D[\mathcal{W}]^{[1]}. 
\end{equation}
The first-order contribution accounting for the difference between the derivatives is given by
\begin{equation}
    \label{S6-eq69}
    D[\mathcal{W}]^{[1]} = \frac{1}{|p_c|}\frac{1}{\tilde{N}}\left[\{\mathcal{W},\tilde{H}_{\text{KS}}[\tilde{N}]\}^{[1]} + \{\mathcal{W},\tilde{H}_{\text{KS}}^{[1]}[\tilde{N}]\} - \left[\left|p_c\right|\tilde{N}\right]^{[1]}\partial_{\eta}\mathcal{W}\right].
\end{equation}
Three distinct contributions can be identified in this expression. The first arises from possible corrections to the Poisson brackets, the second from modifications of the effective Hamiltonian, and the third from the prefactor $|p_c^{\text{eff}}|\tilde{N}^{\text{eff}}$ multiplying the effective Poisson bracket. If a given effective theory does not include one or more of these corrections, the corresponding contributions can simply be omitted. Returning to Eq. \eqref{S6-eq68}, we see that its right-hand side acts as a source term involving known background quantities. As shown in Eq. \eqref{S3-eq21}, these quantities can be expressed, within the original Hamiltonian dynamics, in terms of $\text{X}$ and the three constants of motion of the underformed background model. The left-hand side can be simplified in turn by exploiting the properties of the algebraically special modes given in Eq. \eqref{S4-eq26}. This simplification removes the nonlinearity of the differential equation, reducing the problem to a set of quadratures. We obtain
\begin{equation}
    \label{S6-eq70}
    \partial_{\chi}[\mathfrak{U}^2\mathcal{W}_{\text{ax}}^{[1]} - \mathcal{I}_{\text{ax}}^{[1]}(\text{X})] = 0, \qquad \mathcal{W}_{\text{ax}}^{[1]} = \frac{1}{\mathfrak{U}^2}[\mathcal{I}_{\text{ax}}^{[1]}(\text{X}) + \mathfrak{c}_{\text{ax}}^{[1]}],
\end{equation}
where $\mathfrak{c}_{\text{ax}}^{[1]}$ is an integration constant. The integral is formally given by
\begin{equation}
    \label{S6-eq71}
    \mathcal{I}_{\text{ax}}^{[1]}(\text{X}) = \int^{\text{X}}\left(\mathcal{V}_{\text{ax}}^{[1]} - \mathfrak{b}_{\text{ax}}^{[1]} - D[\mathcal{W}]^{[1]}\right)\frac{\mathfrak{U}^2\text{d}\text{Y}}{\mathbf{\Pi}^2\tilde{\text{F}}(\text{Y})}.
\end{equation}
Before performing the integration, all the background functions in the integrand must be consistently expressed in terms of $\text{X}$.

Having obtained the effective linear correction to the generating function of the Darboux transformation for the axial family, we can proceed to higher orders. In particular, when solving the quadratic equation, the first-order contribution enters the source term alongside the zeroth-order solution. More generally, the nonlinear part of the differential equation retains the same form at every order. Consequently, after minor adjustments to the recurrence relation in Eq. \eqref{S6-eq67}, the solution at order $n$ can be written as
\begin{equation}
    \label{S6-eq72}
    \partial_{\chi}[\mathfrak{U}^2\mathcal{W}_{\text{ax}}^{[n]} - \mathcal{I}_{\text{ax}}^{[n]}(\text{X})] = 0, \qquad \mathcal{W}_{\text{ax}}^{[n]} = \frac{1}{\mathfrak{U}^2}[\mathcal{I}_{\text{ax}}^{[n]}(\text{X}) + \mathfrak{c}_{\text{ax}}^{[n]}],
\end{equation}
where $\mathfrak{c}_{\text{ax}}^{[n]}$ denotes the integration constant at order $n$, and the integral now contains a more complicated source term,
\begin{equation}
    \label{S6-eq73}
    \mathcal{I}_{\text{ax}}^{[n]}(\text{X}) = \int^{\text{X}} \left(\mathcal{V}_{\text{ax}}^{[n]} - \mathfrak{b}_{\text{ax}}^{[n]} - \sum_{m=0}^{n-1}D[\mathcal{W}_{\text{ax}}^{[m]}]^{[n-m]} - \sum_{m=1}^{n-1} \mathcal{W}_{\text{ax}}^{[m]}\mathcal{W}_{\text{ax}}^{[n-m]}\right)\frac{\mathfrak{U}^2\text{d}\text{Y}}{\mathbf{\Pi}^2\tilde{\text{F}}(\text{Y})}.
\end{equation}
Equations \eqref{S6-eq72} and \eqref{S6-eq73} allow us to recover the solution for the linear deformation order as a particular case of the general expression.

At present, two major challenges remain to be addressed. The first is to devise a consistent prescription for fixing the integration constants $\mathfrak{c}_{\text{ax}}^{[n]}$. A natural criterion is to require the correction to the superpotential to vanish at least at one of the two asymptotic ends of the tortoise coordinate in the exterior region. This would ensure that the boundary conditions of the transformed solutions remain unchanged at that end after applying the effective Darboux transformation. The second challenge is to develop a suitable method for evaluating the integrals. Even without specifying the explicit form of the source terms inside the parentheses, the integration is already nontrivial because of the integrating factor $\mathfrak{U}^2/[\mathbf{\Pi}^2\tilde{\text{F}}(\text{Y})]$, which is proportional to the algebraically special modes and has a complicated expression even when evaluated on shell, where the calculation is expected to be simpler. In this regard, progress may be possible by employing numerical integration techniques or approximating the relevant functions with simpler expressions that are more manageable to integration.

\section{Conclusions and outlook}\label{S-VI}

In this work, we have revisited the problem of isospectrality of BH perturbations from a Hamiltonian perspective, focusing on the phase-space structures underlying the relation between axial and polar perturbations. Rather than computing quasinormal-mode spectra explicitly, we have investigated the canonical counterparts of the superpotential, Darboux transformations, and algebraically special modes. We have also explored how the relations connecting these objects are modified in effective theories beyond general relativity. Our analysis provides a unified framework for studying the mathematical structures supporting isospectrality and for characterizing their deviations when the gravitational dynamics is modified.

As a first step, we have introduced an alternative parametrization of the background phase space that makes the connection with the standard Schwarzschild description more transparent. In addition to the BH mass and the background Hamiltonian, we have identified a conserved quantity associated with the asymptotic normalization of the metric. By suitably redefining this quantity, we have obtained a phase-space expression such that its Poisson bracket with the background Hamiltonian vanishes identically. Together with an appropriately defined radial variable, these constants of motion provide a convenient parametrization of phase-space functions and allow the generalized Regge-Wheeler and Zerilli potentials to be expressed in a form closer to their Lagrangian counterparts. This construction facilitates the comparison between the two formulations while retaining the freedom to analyze the background without imposing a particular gauge choice.

Building on this phase-space parametrization, we have investigated the realization in phase space of the structures relating the axial and polar sectors. We have shown that the canonical expressions corresponding to the Regge-Wheeler and Zerilli potentials satisfy the expected superpotential relations on shell, thereby recovering the standard isospectral structure of Schwarzschild perturbations. Off shell, however, there is an obstruction: additional contributions proportional to the background Hamiltonian prevent these relations from holding identically throughout the  background phase space if left unconstrained at zeroth perturbative order. We have traced this discrepancy, at least in part, to a subtle issue, namely, the freedom in the original canonical formulation to absorb terms proportional to the background Hamiltonian in the quadratic perturbative contribution to the global Hamiltonian constraint, by means of a redefinition of the lapse function at our truncation order in the perturbations (which, for the action of the system, is the quadratic order). Although such redefinitions leave the dynamics of the perturbations formally unchanged, they modify the off-shell representatives of the perturbative potentials. It is in this sense that they affect their algebraic relations. This result highlights the importance of distinguishing between the off-shell realization of isospectrality at a certain truncation order in perturbation theory and the physical correspondence between perturbations on solutions of the background equations. 

We have also constructed canonical counterparts of the algebraically special modes. Throughout this work, the term algebraically special has been used repeatedly to refer exclusively to specific properties of the perturbation equations, and should not be confused with the algebraic classification of the background spacetime.  As in the case of the potentials, the expected relations between algebraically special modes and phase-space superpotentials are recovered on shell, whereas their off-shell extensions receive corrections proportional to the background Hamiltonian.

These results complete the phase-space picture of the connection between algebraically special solutions, superpotentials, and Darboux transformations. In particular, they show how the familiar structures of the Lagrangian description can be recovered within a canonical framework in which both the background geometry and the perturbations are treated dynamically. The canonical transformations relating the axial and polar sectors provide a complementary realization of the same structure, preserving the symplectic form and the diagonal form of the perturbative Hamiltonian (at the truncation order of our perturbative analysis).

We have subsequently used this framework to investigate the robustness of isospectrality under effective modifications of the background dynamics. Adopting a prescription in which the functional form of the classical perturbative potentials on the background phase space is retained, while the background variables are replaced by their effective counterparts, we have derived the consistency conditions that must be satisfied for the effective axial and polar equations to be related by a Darboux transformation. These conditions show that the existence of a conserved quantity associated with the superpotential is necessary for preserving the corresponding structure, but is not sufficient: the effective potentials must also satisfy the appropriate differential relations. Since these conditions are not generically preserved by modifications of the background dynamics, isospectrality is expected to be fragile under effective corrections. This conclusion concerns the mathematical structure relating the two master equations; establishing the corresponding quasinormal-mode spectra still requires a separate analysis of the relevant boundary conditions. To characterize departures from the classical relations, we have formulated a perturbative procedure for reconstructing the generating function of an effective Darboux transformation order by order in a deformation parameter associated with the effective description. At each order, the correction satisfies a differential equation with a source determined by the effective modifications, the difference between the effective and classical phase-space derivatives, and the corrections obtained at lower orders. In the linear case, the use of the classical algebraically special modes reduces the problem to quadratures, while the same recursive structure extends formally to higher orders. This provides a systematic way of quantifying the deviations from the classical Darboux relations without requiring the effective axial and polar potentials (obtained with the naive procedure of replacing the background variables with their effective counterparts) to belong to the same Darboux-connected family. The resulting construction is formal at this stage, however, and its application to specific models requires a consistent prescription for fixing integration constants and evaluating the corresponding integrals. In this sense, we have suggested that a reasonable prescription could be the vanishing of the effective corrections to the potential at spatial infinity in the exterior region, while the integrals could be computed in specific cases by means of numerical methods.

Several directions for future research naturally follow from these results, some of which we now comment. First, it would be desirable to revisit the canonical perturbative Hamiltonian without the lapse redefinitions adopted in previous formulations to fix possible contributions of the background Hamiltonian to the perturbative Hamiltonian, and to determine whether alternative definitions of the potentials and of the phase-space derivative yield simpler or more natural off-shell relations. Such an analysis could clarify which aspects of the canonical isospectral structure are independent of the choice of phase-space representatives and which depend on the particular choice. Second, the effective framework developed here should be applied to specific models of modified gravity or effective BH dynamics. This would make it possible to evaluate the correction terms explicitly, determine whether the required conserved quantities exist, and establish under which conditions the classical Darboux structure survives or breaks down. Finally, a natural extension would be to connect the deviations from the Darboux relations derived here with observable differences between the axial and polar quasinormal-mode spectra. This would require solving (analytically, semianalytically, or numerically) the effective master equations with the appropriate boundary conditions and determining whether, and to what extent, the phase-space deviations identified in this work translate into measurable spectral shifts (in fact, some numerical works have started to explore this type of isospectrality violation in the context of loop quantum gravity, see Refs. \cite{GMT,CorralOlmedo}). 

\acknowledgments

We are deeply indebted to Michele Lenzi for many valuable discussions and insightful suggestions throughout this work. G.A.M.M. and A.M.-S. acknowledge support from MCIN/AEI/10.13039/501100011033 (Spanish Ministry of Science and Innovation) and the European Social Fund Plus (ESF+) through Grant No. PID2023-149018NB-C41. C.F.S. acknowledges support from MCIN/AEI/10.13039/501100011033 (Spanish Ministry of Science and Innovation) through Grants Nos. PID2022-137674NB-I00 and PID2025-173920NB-C21, and from AGAUR (Generalitat de Catalunya) through Grant No. 2025-SGR-00191. C.F.S. also acknowledges partial support from the María de Maeztu Unit of Excellence Programme (Grant No. CEX2020-001058-M), funded by MCIN/AEI/10.13039/501100011033 (Spanish Ministry of Science and Innovation). A.M.-S. acknowledges support from the PIPF-2023 fellowship funded by the Comunidad Autónoma de Madrid (Reference No. PIPF-2023/TEC-30167).

\end{document}